\makeatletter
\let\@oddhead\@empty
\let\@evenhead\@empty
\makeatother
 \documentclass[final,5p,times,twocolumn,authoryear]{elsarticle}

\usepackage{textcomp}
\usepackage{amssymb}
\usepackage{lipsum}
\usepackage{graphicx}
\usepackage{natbib}
\usepackage[utf8]{inputenc}
\usepackage{textgreek}

\biboptions{numbers,square,comma}
\begin{document}
\begin{frontmatter}

%% Title, authors and addresses

%% use the tnoteref command within \title for footnotes;
%% use the tnotetext command for theassociated footnote;
%% use the fnref command within \author or \affiliation for footnotes;
%% use the fntext command for theassociated footnote;
%% use the corref command within \author for corresponding author footnotes;
%% use the cortext command for theassociated footnote;
%% use the ead command for the email address,
%% and the form \ead[url] for the home page:
%% \title{Title\tnoteref{label1}}
%% \tnotetext[label1]{}
%% \author{Name\corref{cor1}\fnref{label2}}
%% \ead{email address}
%% \ead[url]{home page}
%% \fntext[label2]{}
%% \cortext[cor1]{}
%% \affiliation{organization={},
%%            addressline={}, 
%%            city={},
%%            postcode={}, 
%%            state={},
%%            country={}}
%% \fntext[label3]{}

\title{Tuning vortex critical velocity in Mo$_2$N thin films via striped magnetic domain configuration}

%% use optional labels to link authors explicitly to addresses:
%% \author[label1,label2]{}
%% \affiliation[label1]{organization={},
%%             addressline={},
%%             city={},
%%             postcode={},
%%             state={},
%%             country={}}
%%
%% \affiliation[label2]{organization={},
%%             addressline={},
%%             city={},
%%             postcode={},
%%             state={},
%%             country={}}

\author[first]{G. Blatter}
\ead{gaston.blatter@ib.edu.ar}

\author[first,second]{M. Sirena}
\author[third]{Yeonkyu Lee}
\author[third]{Jeehoon Kim}
\author[first,second]{N. Haberkorn}
\ead{nhaberk@cab.cnea.gov.ar}

\address[first]{Instituto Balseiro, Universidad Nacional de Cuyo, and Comisión Nacional de Energía Atómica, Av. Bustillo 9500, 8400 San Carlos de Bariloche, Argentina.}%Department and Organization
\address[second]{Comision Nacional de Energia Atomica and Consejo Nacional de Investigaciones Cientificas y Tecnicas, Centro Atomico Bariloche, Av. Bustillo 9500, 8400 San Carlos de Bariloche, Argentina.}
\address[third]{Department of Physics, Pohang University of Science and Technology, Pohang, 37673, South Korea.}

\begin{abstract}
%% Text of abstract
We report on the impact of the magnetic domain stripe configuration on the critical velocity of vortices in superconducting/ferromagnetic bilayers. Using a $23$ nm thick Mo$_2$N film, covered by a $48$ nm FePt layer with tunable nanosized striped domains, we demonstrate that flux instability at low magnetic fields depends on the orientation of the stripes. When the stripes are perpendicular to the applied current and act as vortex guides, the velocity values reach $5$ km/s, duplicating those found when configured parallel to the current, creating winding vortex paths. Our results indicate that vortex critical velocities can be tuned by configuring different domain structures, providing a platform to understand vortex dynamics in superconducting microstrips.

\end{abstract}

\begin{keyword}
%% keywords here, in the form: keyword \sep keyword, up to a maximum of 6 keywords
vortex velocity\sep superconductivity\sep striped domain configuration

%% PACS codes here, in the form: \PACS code \sep code

%% MSC codes here, in the form: \MSC code \sep code
%% or \MSC[2008] code \sep code (2000 is the default)

\end{keyword}

\end{frontmatter}

%\tableofcontents

%% \linenumbers

%% main text

\section{Introduction}
\label{introduction}

The ultra-rapid dynamics exhibited by superconducting vortices during dissipation involve complex physics that applies to systems not in equilibrium \cite{PhysRevB.83.144526,PhysRevLett.106.037001}. The maximum attainable vortex velocity in superconducting microstrips, as determined by current-voltage (I-V) curves, is constrained by Larkin-Ovchinnikov (LO) instability \cite{1975JETP...41..960L}. This phenomenon manifests as an abrupt transition to the normal state during dissipation, associating the vortex velocity at the instability ($v_{LO}$) with the time of recombination of normal electrons into Cooper pairs ($\tau$). The LO instability, originally anticipated to be solely governed by intrinsic properties, is also observed to be influenced by local heating generated by irregularities and geometric defects. Reducing surface roughness and minimizing local heating effects, the maximum $v_{LO}$, typically observed at small magnetic fields, can increase from values near $1$ km/s to exceed $10$ km/s \cite{PhysRevB.95.184517, PhysRevB.99.174518,articleUltraFastVortexMotion}. The quest for achieving high vortex velocities while mitigating unintended effects is motivated by two primary reasons. Firstly, accurately determining the $\tau$ through LO theory is crucial. This is particularly relevant because $\tau$ dictates the maximum resolution achievable when utilizing the material in superconducting nanowire single-photon detectors (SNSPD) \cite{Natarajan_2012}. Secondly, this pursuit is driven by exploring new phenomena associated with the Cherenkov-like generation of sound and spin waves induced by fast-moving vortices in superconducting/magnetic systems \cite{PhysRevLett.106.037001, PhysRevB.89.054516}. Hence, material engineering, encompassing both intrinsic properties and structural/geometrical characteristics, is crucial in enhancing vortex speeds in micro and nanosystems.

Proximity effects significantly influence the properties of superconducting/ferromagnetic hybrids. Studies concerning vortex dynamics span from pinning mechanisms to the impact on vortex critical velocities at the LO instability \cite{Hoffman_Prieto_Pedro, PhysRevB.84.054536,PhysRevB.96.174519,PhysRevApplied.11.054064}. The proximity to ferromagnetic materials typically enhances vortex velocities and correlates with a shorter $\tau$ (since $v_{LO}^2\propto 1/\tau$) \cite{PhysRevB.84.054536,PhysRevB.96.174519}. Furthermore, domain boundaries within ferromagnetic materials serve as effective guides for vortices, further influencing their behavior. Additionally, it is worth noting that magnetic domain lines play a crucial role in enhancing vortex velocities when compared to single films and superconducting bilayers \cite{PhysRevApplied.11.054064}. An interesting observation is the increase in vortex velocities at the crossover between magnetization in-plane and striped domain structures in Nb/permalloy bilayers \cite{PhysRevB.96.174519}. This observation suggests the possibility of tuning the LO instability in magnetic materials by modifying the domain structure in a same system.

In this study, we investigate the control of vortex critical velocities in superconducting/ferromagnetic hybrids by manipulating the stripe domain configuration. This manipulation directly affects the motion of vortices, which is influenced by the Lorentz force. Our bilayer consists of a $23$ nm thick Mo$_2$N film and a $48$ nm thick FePt film as the ferromagnetic system. Mo$_2$N thin films exhibit a critical temperature ($T_c$) of approximately $8$ K when grown at room temperature \cite{HABERKORN201815}, while FePt films exceeding $40$ nm in thickness undergo a transition from in-plane magnetization to a striped domain structure \cite{10.1063/1.4942652}. Moreover, upon saturating and subsequently removing the magnetic field, the stripes align themselves parallel to the field direction \cite{PhysRevB.82.144410}. We compare the results obtained when the stripes function as guides for vortices (perpendicular to the current) and as barriers to vortex motion (parallel to the current). While a previous study \cite{PhysRevB.96.174519} demonstrated the effectiveness of stripe domains in increasing vortex velocities, our work takes a step further by not only confirming their effectiveness but also showcasing the impact of modifying their configuration to fine-tune critical vortex speeds during dissipation. Additionally, our findings contribute to a deeper understanding of the role of disorder and vortex path winding in the LO instability.

\begin{figure}[ht]  % [htb] son opciones de ubicación: aquí, parte superior o inferior
  \centering
  \includegraphics[width=1\linewidth]{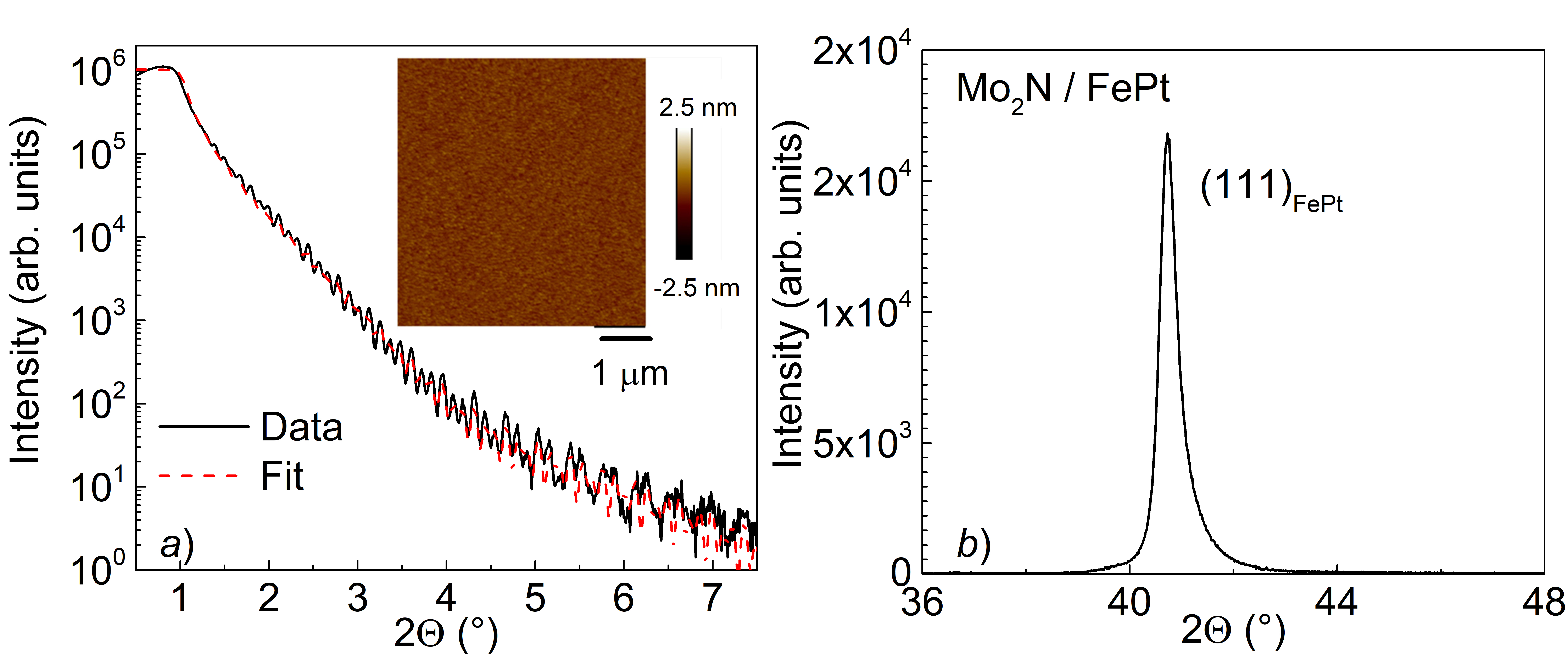}
  \caption{Mo$_2$N ($23$ nm) / FePt bilayer ($48$ nm). a) Low angle XRD. Inset: AFM image for the surface of the sample. b) XRD pattern.}
  \label{fig1}
\end{figure}

\section{Methods}
%%\label{Mehods}
A Mo$_2$N / FePt bilayer was grown through reactive sputtering at room temperature on a (100) Si substrate. The base pressure in the chamber was $4\times10^{-5}$ Pa. The Mo$_2$N layer grew in a $6\%$ N$_2$ atmosphere at $0.66$ Pa (N$_2$$+$Ar), positioned above the Mo target ($100$ W) at $0.15$ m. The FePt layer was grown in pure argon at $0.4$ Pa using $20$ W at $0.1$ m target distance. The bilayer consisted of a $23$ nm thick Mo$_2$N and $48$ nm of FePt. The thickness for Mo$_{2}$N was selected considering that $T_c$ is weakly affected by dimensional effects \cite{HABERKORN2022139475}.

XRD data (\texttheta \ - 2\texttheta   
\ configuration) were obtained using Panalytical Empyrean equipment at $40$ kV and $30$ mA with CuK$_\alpha$ radiation and an angular resolution of $0.013 ^o$. Atomic force (AFM) and magnetic force microscope (MFM) measurements were conducted on a Dimension $3100$ \textcopyright Brucker microscope in tapping mode. Electrical transport measurements were performed on an $80$  (L) $\times$ $5$ $\mu$m (w) bridge using standard four-terminal transport. Magnetic measurements were done with a commercial SQUID. Bridges were fabricated using optical lithography and argon ion milling. Current-voltage (IV) curves were obtained with a Keithley Nanovoltmeter Model 2128A and a Keithley Current source Model 6221 AC/DC in synchronized mode with a $0.1$ ms pulse duration.

\section{Results and discussion}
%%\label{Results and discussion}

Figure \ref{fig1} shows low-angle diffraction data for the studied sample and the corresponding fitting using the Parrat code \cite{PARRAT}. The pattern displays well-defined maxima and minima, which are characteristic of samples with low roughness. The thicknesses obtained from the fit correspond to $23$ nm for Mo$_2$N and $48$ nm for FePt. The flatness of the sample's surface was confirmed by AFM (see inset in Figure \ref{fig1}a), which appeared to be free of defects and exhibited a root mean square (RMS) roughness of $0.2$ nm. Figure \ref{fig1}$b$ displays an XRD pattern for the bilayer, where the (111) peak corresponds to a textured, disordered face-centered cubic (FCC) structure \cite{10.1063/1.4942652}. The reflection (200), which is usually observed at approximately $43^o$ for $\gamma$-Mo$_2$N, does not appear due to the nanocrystalline nature and thickness of the film \cite{HABERKORN201815}. Furthermore, the latter peak is masked by the stronger reflection from the FePt.

\begin{figure}[ht!]  % [htb] son opciones de ubicación: aquí, parte superior o inferior
  \centering
  \includegraphics[width=\linewidth]{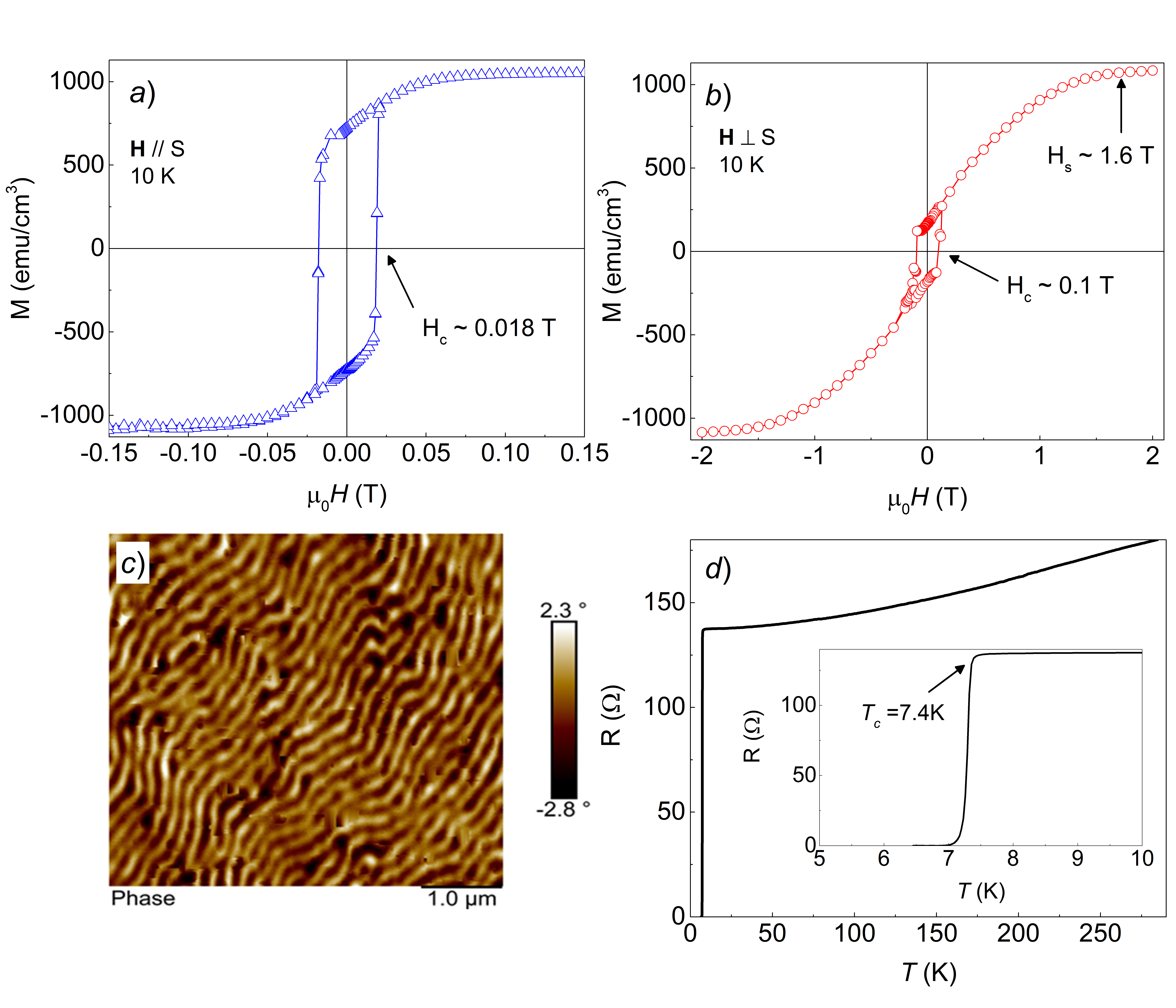}
  \caption{Mo$_2$N ($23$ nm) / FePt bilayer ($48$ nm). a-b) Magnetic hysteresis loops at 10 K with the field applied parallel and perpendicular to the sample surface. c) MFM image of the magnetic domain structure at room temperature. d) Temperature dependence of the electrical resistance without applied field. Inset: zoom of the superconducting transition. }
  \label{fig2}
\end{figure}

Figures \ref{fig2}$ab$ compare the magnetic hysteresis loops at $10$ K, with the magnetic field applied parallel (\textbf{H}//S) and perpendicular to the surface (\textbf{H}$\perp$S). The saturation magnetization for the sample is approximately $1100$ emu/cm$^3$, which aligns with the expected values for the material \cite{10.1063/1.4942652,PhysRevB.66.024413}. The hysteresis loop for \textbf{H}//S displays typical features of stripe-like domains, including significant coercivity ($H_c$ $\approx 0.018$ T) and a reduction in remanence (compared to saturation) due to the perpendicular component of the stripes. The hysteresis loops with \textbf{H}$\perp$S show remnant magnetization attributed to the out-of-plane component produced by the stripes, with $H_c$ $\sim 0.1$ T. Additionally, saturation dominated by shape anisotropy occurs at $H_s$ $\approx 1.6$ T. MFM images at room temperature (Figure \ref{fig2}$c$) confirm the presence of magnetic stripes, with a width of approximately $40$ nm, a value relevant for comparing with the distance between vortices fixed by the magnetic field. Indeed, the distance between maxima and minima in the field modulation produced by the stripes is of $\approx 80$ nm, which assuming that it does not change with temperature, corresponds to an intervortex distance $a=1.073$ with $\mu$$_0H$ $\approx 0.37$ T. In addition to magnetic characterization, we determined the sample's $T_c$ by measuring resistance as a function of temperature. The bilayer exhibits metallic behavior due to FePt, with a $T_c$ of $7.4$ K (Figure \ref{fig2}$d$, main panel, and inset), in accordance with expectations for single films of similar thickness \cite{HABERKORN2022139475}.

\begin{figure*}[ht!]  % [htb] son opciones de ubicación: aquí, parte superior o inferior
  \centering
  \includegraphics[width=\linewidth]{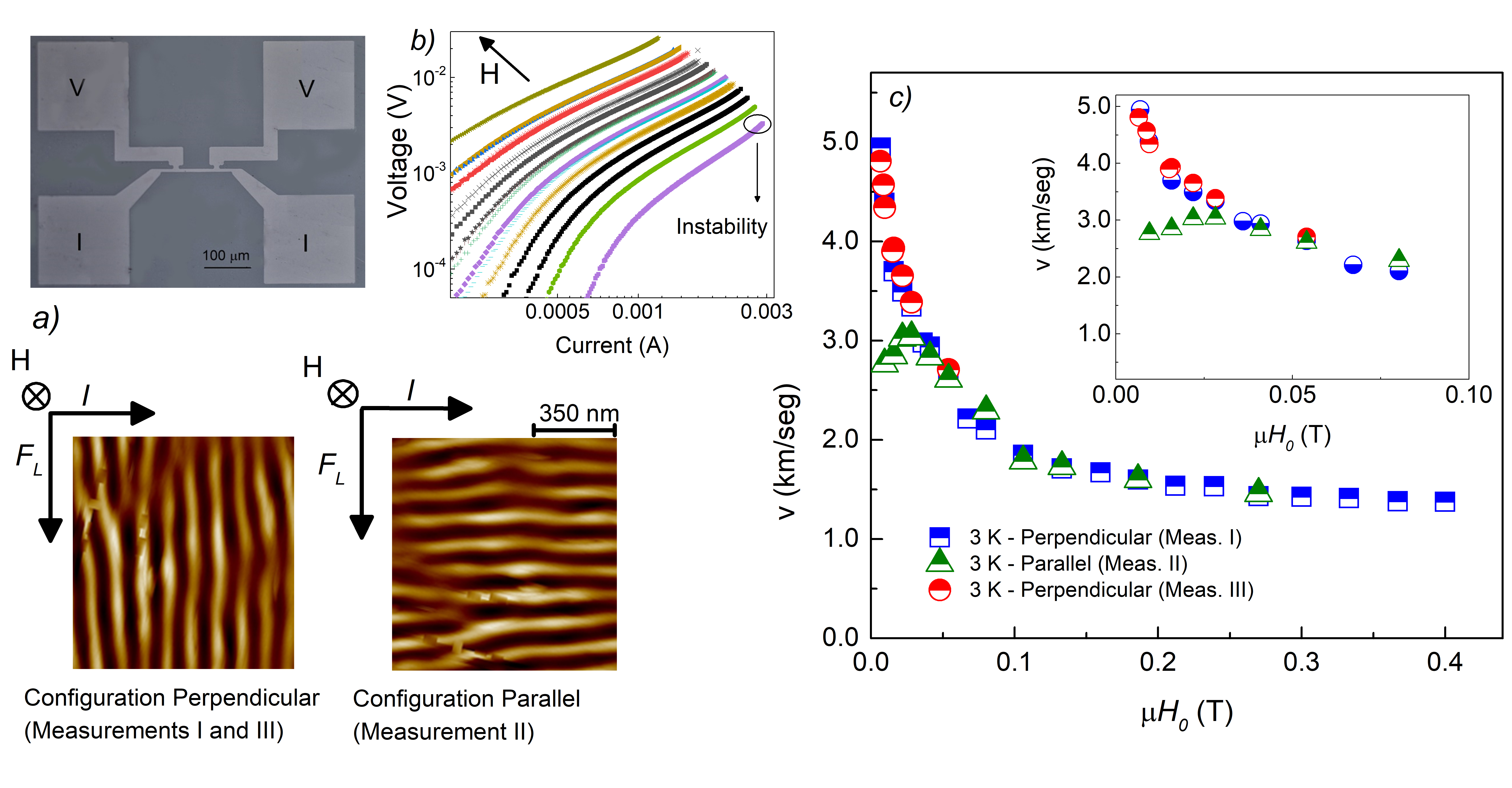}
  \caption{a) Electrical bridge and configurations between the magnetic domains structure the electrical current. b) Typical IV curves and the criterion for the LO instability. c) Magnetic field dependence of the critical vortex velocity for different magnetic domain configurations. Inset: zoom of the low field range }
  \label{fig3}
\end{figure*}

We investigated the influence of the magnetic domain configuration on vortex instability. To make the comparison, we aligned the magnetic stripe domains either perpendicular or parallel to the applied current in the microbridge (as shown in Figure \ref{fig3}$a$). We achieved this by rotating the sample and the magnetic field orientation, either in-plane (to fix the domains) or out-of-plane for IV curves. Initially, we saturated the sample with \textbf{H}//S, removed the field, and then rotated the sample to apply \textbf{H}$\perp$S. We performed measurements for each configuration during different cooldowns labeled as measurement I (parallel), II (perpendicular), and III (parallel). The analysis was conducted at $3$ K, measuring IV curves as a function of the magnetic field (see typical curves in Figure \ref{fig3}$b$). The vortex velocity can be obtained as $v= \frac{V}{\mu_0HL}$, where $V$ is the voltage at the instability and $L$ is the distance between voltage contacts. Measurements I and III (Figure \ref{fig3}c) were performed to show that the process is reproducible in these extreme configurations. Our results indicate that the magnetic field configuration influences vortex velocities at low fields ($\mu_0H< 0.05$ T). Specifically, when the stripes act as vortex guides (measurements I and III), velocities reach $\approx 5$ km/s at low field and it exhibits the typical decay with field commonly observed in $v_{LO} (H)$ dependencies \cite{PhysRevApplied.11.054064}. In contrast, velocities decrease significantly and deviate from the typical decay with the field for $\mu_0H < 0.05$ T when stripes create winding vortex paths (measurement II). Both magnetic configurations exhibit similar behavior for magnetic fields exceeding $0.05$ T, indicating that as the magnetic field increases, the magnetic paths tend to vanish. This value is much smaller than the $\mu_0H = 0.37$ T estimated as the optimal inter-vortex distance determined by the stripe width. It is important to mention that despite differences at low magnetic field, the data show a $v_{LO}$$\propto$ $H^{-0.5}$ dependence, as is expected for finite heat removal from the substrate \cite{BEZUGLYJ1992234}. Since the quasiparticle diffusion constant $D$ may be influenced by proximity to a conducting metal \cite{Ustavschikov_2021}, we have chosen to refrain from extracting $\tau$ using the A. Bezuglyj and V. Shklovskij model \cite{BEZUGLYJ1992234}.

While previous studies demonstrated the effectiveness of stripe domains and vortex guides in increasing vortex velocities \cite{PhysRevB.96.174519,PhysRevApplied.11.054064}, our work not only reaffirms the importance of vortex configuration but also highlights the impact of modifying this configuration to fine-tune critical vortex speeds during dissipation, transitioning from vortex guides to winding paths. In addition to analyzing the influence of the striped domain configuration on vortex critical velocities, it is worthwhile to conduct a detailed examination of the impact of the proximity of the Mo$_2$N layer to a ferromagnetic material. When compared to a Mo$_2$N film of similar thickness, we observed that velocities at low magnetic fields increased from approximately $3$ km/s to approximately $5$ km/s \cite{HABERKORN2022139475}. Furthermore, the values at moderate magnetic fields increased significantly, reaching approximately $1.5$ km/s at $\mu_0H = 0.4$ T, compared to approximately $0.7$ km/s for a Mo$_2$N single layer. This increase may be attributed to the ferromagnetic nature of the FePt layer as well as its thermal conductivity due to its metallic properties. As we recently reported, for moderate and high magnetic field conditions, superconducting/metal bilayers exhibit an increase in vortex velocities that scale with the thickness and thermal conductivity of the metal used \cite{BLATTER2024116943}. In the case of the FePt layer, this effect may be associated with heat dissipation and the potential contribution of its proximity to the ferromagnetic material. On the other hand, higher vortex velocities can be analyzed within two different scenarios: 1) smaller $\tau$ \cite{PhysRevB.84.054536,PhysRevB.96.174519}, or 2) small non-equilibrium effects related to Joule heating impacting the instability of vortex motion \cite{Ustavschikov_2021}. It is important to note that vortex velocities at low magnetic fields of approximately $5$ km/s are similar to those reported in other superconducting/ferromagnetic systems \cite{PhysRevB.84.054536,PhysRevB.96.174519,PhysRevApplied.11.054064}, which suggests a common mechanism leading to an increase in vortex velocities. Finally, concerning the winding paths of vortices, our results unequivocally show that irregularities within the superconducting material can lead to a decrease in vortex velocities, a phenomenon not solely attributable to the constraints imposed by intrinsic superconducting properties in the LO theory. Introducing tunable magnetic domains may offer a promising avenue to enhance vortex velocities and investigate the physics associated with the interaction between magnetic moments in superconducting/ferromagnetic hybrids.

\section{Conclusions}
%%\label{Conclusions}

In summary, we examined the potential to adjust vortex velocities in superconducting/ferromagnetic bilayers by modifying the nanoscale magnetic domain configuration. Our findings contrast two extreme configurations of stripes, one providing magnetic guides for rapid motion and the other creating winding vortex paths for slower motion. Although the effect is primarily noticeable at low magnetic fields, where the magnetic modulation caused by stripe domains is expected to be minimally influenced by external fields, our results using FePt with nanosized stripe domains offer a straightforward platform for investigating vortex dynamics and critical vortex behavior in superconducting microsystems.

\section*{Acknowledgements}

This work was partially supported by the ANPCYT (PICT 2018- 01597), U. N. de Cuyo 06/C013T1, CONICET (PIP 11220210100263CO), BrainLink program funded by the Ministry of Science and ICT through the National Research Foundation of Korea (2022H1D3A3A01077468) and Brain Pool program funded by the Ministry of Science and ICT through the National Research Foundation of Korea (RS-2023-00222408). JK and JY were supported by the National Research Foundation of Korea (NRF) grant funded by the Korean government (MSIT) (Grant No. NRF-2019R1A2C2090356) and the Technology Development Program (Grant No. S3198743) funded by the Ministry of SMEs and Startups (MSS, Korea). MS and NH are members of the Instituto de Nanociencia y Nanotecnología INN (CNEA-CONICET). 

\section*{Author contribution}

GB and NH were responsible for growing the samples and conducting XRD analysis, as well as performing electrical transport measurements. MS conducted AFM and MFM measurements and analysis. GB and NH wrote the manuscript, while all authors contributed to the data discussion.

\section*{Declaration of competing interest}

The authors declare that they have no known competing financial interests.

\bibliography{example}  % Specify the name of your .bib file here

%% If you have bibdatabase file and want bibtex to generate the
%% bibitems, please use
%%

%% else use the following coding to input the bibitems directly in the
%% TeX file.

%%\begin{thebibliography}{00}

%% \bibitem[Author(year)]{label}
%% For example:https://www.overleaf.com/project/652230c0f8ffa95c6c0eddbb

%% \bibitem[Aladro et al.(2015)]{Aladro15} Aladro, R., Martín, S., Riquelme, D., et al. 2015, \aas, 579, A101

%%\end{thebibliography}

\end{document}